\documentclass[12pt,a4paper]{iopart}
\usepackage{iopams} 

\usepackage[colorlinks,backref]{hyperref}
\usepackage{%
  units,%
  graphicx,%
  color}

\renewcommand{\vec}[1]{\ensuremath{\boldsymbol{#1}}}
\newcommand{\ket}[1]{\ensuremath{\left|{#1}\right>}}
\newcommand{\bra}[1]{\ensuremath{\left<{#1}\right|}}
\newcommand{\gstate}{\ensuremath{\ket{\textnormal{g}}}}
\newcommand{\estate}{\ensuremath{\ket{\textnormal{e}}}}
\newcommand{\rstate}{\ensuremath{\ket{\textnormal{r}}}}
\newcommand{\rcstate}{\ensuremath{\bra{\textnormal{r}}}}
\newcommand{\fiveS}{\ensuremath{\textnormal{5S}_{1/2}}}
\newcommand{\fiveP}{\ensuremath{\textnormal{5P}_{3/2}}}
\newcommand{\ftS}{\ensuremath{\textnormal{43S}_{1/2}}}
\newcommand{\twopit}{\ensuremath{2\pi\times}}
\newcommand{\subt}[2]{{#1}_{\textnormal{\footnotesize{#2}}}}
\newcommand{\eqref}[1]{(\ref{#1})}
\definecolor{brown}{rgb}{0.73,0.51,0.36}
\definecolor{blue}{rgb}{0.118,0.102,0.420}

\begin{document}

\title{Investigation of dephasing rates in an interacting Rydberg gas}

\author{U~Raitzsch$^1$, R~Heidemann$^1$, H~Weimer$^2$, V~Bendkowsky$^1$, B~Butscher$^1$, P~Kollmann$^1$, R~L\"{o}w$^1$, H~P~B\"{u}chler$^2$ and T~Pfau$^1$}

\address{$^1$ 5. Physikalisches Institut \\ 
$^2$ Institut f\"{u}r Theoretisches Physik III \\
Universit\"{a}t Stuttgart, Pfaffenwaldring 57, 70550 Stuttgart, Germany}
\ead{t.pfau@physik.uni-stuttgart.de}
\begin{abstract}
 We experimentally and theoretically investigate the dephasing rates of the coherent evolution of a resonantly driven pseudo spin emersed in a reservoir of pseudo spins. The pseudo spin is realized by optically exciting $^{87}$Rb atoms to a Rydberg state. Hence, the upper spin states are coupled via the strong van der Waals interaction. Two different experimental techniques to measure the dephasing rates are shown: the `rotary echo' technique known from nuclear magnetic resonance physics and electromagnetically induced transparency. The experiments are performed in a dense frozen Rydberg gas, either confined in a magnetic trap or in an optical dipole trap. Additionally, a numerical simulation is used to analyse the dephasing in the rotary echo experiments.

\end{abstract}

% Keywords Required only for MST, PB, PMB, PM, JOA, JOB? 
% \submitto{\NJP}
\tableofcontents
\maketitle

%\doublespacing
\section{Introduction}
The detailed understanding of dephasing processes in a quantum system is essential if one desires to obtain coherent control over quantum matter. Spin-$\nicefrac{1}{2}$ systems imbedded into a strongly coupled environment are the paradigm for decoherence theory. Most noticeable the spin-boson model has been studied in great depth \cite{Leg87,Weiss08}. Ultracold atoms, that are coherently excited into a Rydberg state, can serve as a test bed for strongly interacting quantum matter. The coherent dynamics of an individual spin is coupled strongly to the bath of the other simultaneously driven atoms leading to a dephasing of the quantum mechanical evolution. In this paper the dephasing mechanism is investigated experimentally by two complementary methods and compared to a numerical model calculation. This work is also related to possible quantum computational schemes as frozen Rydberg gases, i.e. the centre of mass motion of the atoms is negligible on the time scale of the experiment, can be used to build fast quantum gates \cite{Jak00,Luk01}. The latter article suggests to use the strong interaction between Rydberg atoms, namely the dipole-dipole interaction, to create a mesoscopic collective state, which has the advantage that not every single atom must be controlled separately on a microscopic scale. The blockade of the excitation into the Rydberg state due to the dipole-dipole and van der Waals interaction has been observed in laser cooled atomic clouds in various experiments \cite{Sin04,Ton04,Cub05a,Vog06} and in magnetically trapped atomic samples at much higher number densities \cite{Hei07}. 

The coherent excitation of interacting Rydberg systems is investigated in \cite{Hei07,Moh07,Rai08,Wea08}. In order to successfully build a quantum gate operation it is crucial to know the mechanisms and the time scales on which the dephasing of the states happen in the system. This knowledge can be incorporated into the technical realization of quantum gate operation in future experiments. The gate operation in a quantum system must take place on a time scale faster than the dephasing rate $\subt{\gamma}{d}$ in order to avoid decoherence of the system. In this letter the dephasing rates are investigated by two complementary methods, namely using the rotary echo technique \cite{Sol59} and electromagnetically induced transparency \cite{Har90,Bol91}, are investigated. 

Further insight in the processes causing the dephasing is obtained by numerical simulations of the rotary echo experiment. The rotary echo technique is known from the research field of nuclear magnetic resonance and provides a tool to overcome disturbing effects due to inhomogeneous distributed Rabi frequencies. An echo proves the coherence of a system directly by exciting and deexciting the atomic sample by a time reversal in the driving field. 

Another way to investigate the coherence properties of the Rydberg sample is the observation of electromagnetically induced transparency. Here the coherence of the system is proven by the existence of a coherent superposition between two states, namely the ground state and the Rydberg state.

\section{Setup}

\begin{figure}[ht]
  \includegraphics{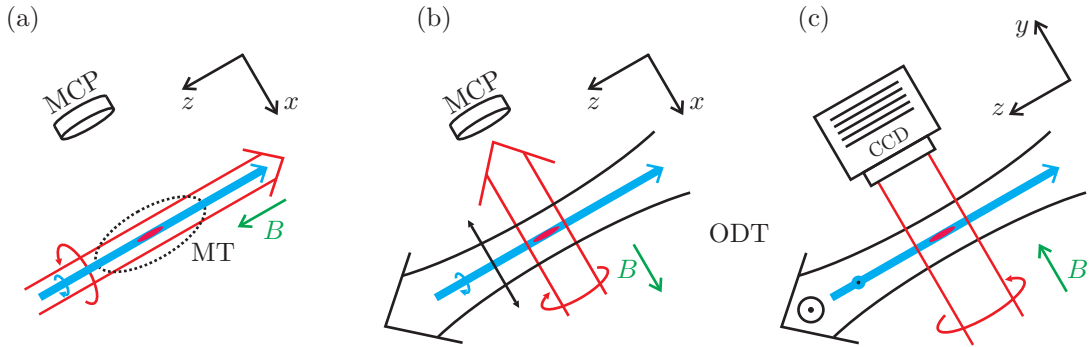}
  \caption{Schematic view of the experimental setup used for the rotary echo experiment in the magnetic trap (a) and the dipole trap (ODT) (b). Figure (c) shows the setup for the EIT experiment in an ODT. The ODT beam is depicted in black and propagating along the $z$-direction. The laser is linearly polarized along the $x$-axis and far red detuned (\unit[826]{nm}) with respect to the ground state. For the rotary echo sequence in the magnetic trap (dashed ellipse) both excitation lasers are collinear and propagating along the -$z$-direction. The laser for the lower transition (see \fref{fig:rotesEIT} (a)) of the two-photon excitation into the \ftS\, via the \fiveP , shown in red, has a wavelength of \unit[780]{nm} and a diameter of $~\unit[1]{mm}$. The laser for the excitation into the \ftS\, Rydberg state is shown in blue and has a wavelength of \unit[480]{nm} and a waist of $\unit[42]{\mu m}$. The lasers are $\sigma^+$ (\unit[780]{nm}) and $\sigma^-$ (\unit[480]{nm}) polarized with respect to the quantization axis (green arrow) given by a small magnetic field along the $z$-direction. In the ODT the quantization is chosen along $x$, while the \unit[780]{nm} beam is traveling along -$x$. For the EIT sequence a quantization axis along the $y$-direction is chosen. The \unit[780]{nm} beam is $\sigma^+$ polarized and imaged with a CCD camera, which is also used in the rotary echo sequence to image the ground state atoms after the Rydberg excitation. The Rydberg atoms are field ionized and detected by a multichannel plate.}
 \label{fig:Setup}
\end{figure}

In this letter two different experimental schemes are presented for the investigation of the dephasing of the driven Rydberg system excited from a cloud of ultracold $^{87}$Rb atoms, trapped either in a magnetic trap or an optical dipole trap in the \fiveS$(f=2,m_f=2)$ state. The first experiments discussed here are rotary echo experiments \cite{Sol59,Rai08} in a magnetic trap and in an optical dipole trap. The second experiment uses electromagnetically induced transparency (EIT) of an atomic sample trapped in an optical dipole trap. Details of the experimental setup are extensively discussed in \cite{Loe07}. \Fref{fig:Setup} shows specific information about the configurations used in the different experiments. 

\subsection{Magnetic trap}\label{subsection:MagneticTrap}

Each experimental sequence described in this paper starts with an evaporatively cooled cloud of $^{87}$Rb atoms in a magnetic trap. The trapping frequencies are radially $\omega_r=\unit[\twopit 322]{Hz}$ and axially $\omega_z=\unit[\twopit 18]{Hz}$. In order to study the density dependence of the dephasing the density of ground state atoms in the trap is varied by means of a Landau-Zener sweep \cite{Zen32}. Using this technique atoms are transferred from the initial \fiveS$(f=2,m_f=+2)$ of $^{87}$Rb via a microwave photon with a frequency of \unit[6.8]{GHz} into the magnetically untrapped \fiveS$(f=1,m_f=+1)$ state. This is the starting point for the rotary echo measurement done with magnetically trapped atoms described in section \ref{subsection:RE}. 

\subsection{Optical dipole trap}\label{subsection:OpticalDipoleTrap}

An optical dipole trap (ODT) was set up to measure the dephasing rates using the rotary echo sequence (see section \ref{subsection:RE}) and the electromagnetically induced transparency with the advantage that a homogenous magnetic field over the atomic sample can be applied in arbitrary direction. Regarding the experiments on EIT (see section \ref{section:EIT}) it is important to switch off the trapping potential quickly. This is much easier to achieve in an ODT than in a magnetic trap. Furthermore, it is possible to tailor the geometry of the cloud in a wide range using optical potentials. The laser beam for the ODT is propagating collinearly to the long axis of the cigar-shaped cloud in the $z$-direction and is linearly polarized along $x$ (see \fref{fig:Setup}). It has a wavelength of \unit[826]{nm}, a waist of $w_0 = \unit[21]{\mu m}$ and a quality factor of the Gaussian beam of $M^2=1.5$. The power of the ODT is stabilized using a photo diode and an acousto-optical modulator to $P=\unit[22]{mW}$. With a trap depth $U_0 = \unit[24]{\mu K}$ one ends up with trap frequencies of $\subt{\nu}{r}=\unit[735]{Hz}$ and $\subt{\nu}{z}=\unit[10]{Hz}$ in the radial and axial direction, respectively. 

After evaporative cooling and the Landau-Zener sweep in the magnetic trap the ODT is ramped up within \unit[50]{ms}. Afterwards the magnetic trap is switched off within \unit[20]{ms} and the cloud is kept for \unit[100]{ms} in the ODT for thermalization. 

At a temperature of $\unit[6]{\mu K}$ the atomic cloud has a Gaussian width ($1/\sqrt{e}$-radius) in axial direction of $\sigma_z=\unit[(400\pm 10)]{\mu m}$ and in radial direction of $\sigma_r\simeq\unit[5]{\mu m}$. Due to the spatial resolution of $\unit[5.6]{\mu m}$ of the imaging system it is not possible to accurately resolve the radial size of the atomic cloud.

The peak density of ground state atoms in the harmonic approximation is $\subt{n}{g} = N_0/((2\pi)^{3/2}\,\sigma_r^2\sigma_z)$, with the trapped atom number $N_0$. In order to calculate $\subt{n}{g}$ the atom number $N_0$ and the size in axial direction $\sigma_z$ are taken from absorption images of the atomic cloud. The radial size of the atomic sample is given by $\sigma_r = \omega_r^{-1}(\subt{k}{B} T/m)^{1/2}$, with $\omega_r=\twopit\nu_r$. 

\section{Dephasing measurements}\label{section:DephasingMeasurement}

\subsection{The rotary echo experiment}\label{subsection:RE}

The rotary echo experiment in the magnetic trap is done at a temperature of $T=\unit[3.8]{\mu K}$ with a peak density of ground state atoms of $\subt{n}{g}=\unit[5.2\times 10^{19}]{m^{-3}}$ \cite{Rai08}. The atoms are excited for a time $200\leq\tau\leq\unit[800]{ns}$ into the \ftS\, Rydberg state, by means of a two-photon excitation (see \fref{fig:rotesEIT} (a)). The lasers are detuned with respect to the \fiveP\, state by $\unit[\twopit 470]{MHz}$, but tuned to resonance with respect to the two-photon excitation. As the detuning to the radiative state is large, the system reduces to an effective two-level atom. The lasers propagate along the $-z$-axis with $\sigma^+$ (\unit[780]{nm}) and $\sigma^-$ (\unit[480]{nm}) polarization resulting in a selective excitation of the $\ftS(j=1/2,m_j=+1/2)$ state. This Rydberg state is insensitive to magnetic field inhomogeneities. The waist of the \unit[780]{nm} laser is \unit[1]{mm} and \unit[42]{$\mu$m} for the \unit[480]{nm} laser. 

\begin{figure}[ht]
  \includegraphics{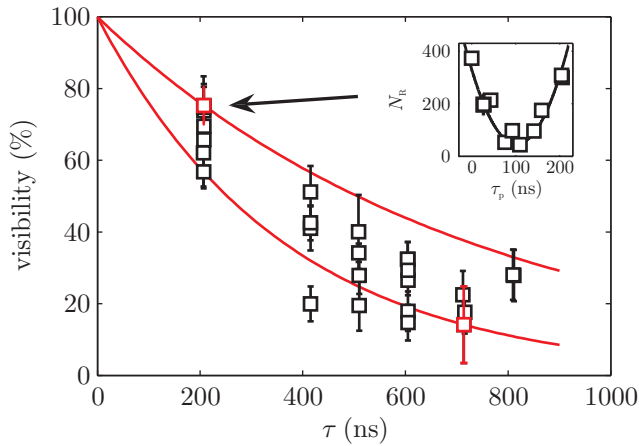}
  \caption{Visibility of the rotary echo signal in the ODT as a function of pulse duration $\tau$ for different densities $\subt{n}{g}$ of ground state atoms. In order to obtain the dephasing rates $\subt{\gamma}{d}=-\ln(V)/\tau$ an exponential decay of the visibility with increasing pulse length is assumed. Two fits are shown exemplarily in red. The inset shows the Rydberg atom number $\subt{N}{R}$ as a function of $\subt{\tau}{p}$, resulting in a typical rotary echo signal. This particular measurement is done for $\tau=\unit[200]{ns}$ and $\subt{n}{g}=\unit[4.4\times 10^{17}]{m^{-3}}$. A visibility of $\unit[75]{\%}$ is obtained from a parabolic fit.}
 \label{fig:expDecay}
\end{figure}

After a time $\subt{\tau}{p}\leq\tau$ the phase of the upper excitation laser is flipped by $\pi$ and thereby also the sign of the effective two-level excitation is inverted from $\Omega$ to $-\Omega$. In contrast, the sign of the interaction among the Rydberg atoms is not changed. Assuming undamped Rabi oscillations without spontaneous decay, interaction or finite linewidth of the excitation laser, the population is completely reversed from the Rydberg state into the ground state when $\subt{\tau}{p}=\tau/2$  (see inset of \fref{fig:expDecay}). The number of atoms in the Rydberg state is measured by field ionization of the Rydberg atoms. Due to the applied electric field the ions are additionally pushed towards a multichannel plate (MCP). The atoms remaining in the ground state are imaged using resonant light on the \fiveS$(f=2,m_f=+2)$ to \fiveP$(f=3,m_f=+3)$ transition in $y$-direction by a charge-coupled device (CCD) camera.

The rotary echo is additionally performed in an optical dipole trap with a slightly different setup compared to the rotary echo experiment of magnetically trapped atoms described in \cite{Rai08}. The \unit[780]{nm} laser for the lower transition has a waist of $\unit[1]{mm}$ and is travelling along the -$x$-direction (see \fref{fig:Setup} (b)). In this scheme the atoms are quantized along $x$. The atoms are radially irradiated to reduce effects due to absorption, i.e. inhomogeneous intensities. The experiment is now performed for densities $\subt{n}{g}$ of ground state atoms between $\unit[3\times 10^{17}]{m^{-3}}$ and $\unit[2\times 10^{18}]{m^{-3}}$.

\Fref{fig:expDecay} shows the visibility of the rotary echo signal in the ODT for various pulse durations $\tau$ and peak densities of ground state atoms $\subt{n}{g}$. The inset of \fref{fig:expDecay} shows a typical rotary echo measurement for a pulse duration of \unit[200]{ns}. The visibility
\begin{eqnarray}
 \label{eq:visibility}
V = \frac{\subt{N}{R}(\subt{\tau}{p}=0)-\subt{N}{R}(\subt{\tau}{p}=\tau/2)}{\subt{N}{R}(\subt{\tau}{p}=0)+\subt{N}{R}(\subt{\tau}{p}=\tau/2)}
\end{eqnarray}
is obtained using a parabolic fit. The dephasing rate $\subt{\gamma}{d}=-\ln(V)/\tau$ is obtained by assuming an exponential decay of the visibility with increasing pulse duration $\tau$. 

In order to study the dephasing of the Rydberg state theoretically numerical simulations of the rotary echo are performed. Therefore the time evolution of a system consisting of $N$ atoms in a box of volume $V$, according to the Hamiltonian
\begin{eqnarray}
 \label{eq:HendrikHampOp} 
 \mathcal{H} = \frac{\hbar\Omega(t)}{2} \sum\limits_i \vec{\sigma}_x^{(i)} + C_{6} \sum\limits_{j<i} \frac{\subt{\hat{P}}{rr}^{(i)}\subt{\hat{P}}{rr}^{(j)}}{|{\vec{r}}_i-{\vec{r}}_j|^6}\, ,
\end{eqnarray}
is computed. Here $\vec{\sigma}_{\alpha}^{(i)}$ are the Pauli matrices, $\subt{\hat{P}}{rr}^{(i)}=\rstate_i\rcstate_i= (1+\vec{\sigma}_z^{(i)})/2$ is the projector onto the excited Rydberg state, $C_6$ denotes the strength of the van der Waals interaction, and the positions of the atoms $\vec{r}_i$ are randomly distributed but fixed. As in the experiment, the Rabi frequency $\Omega(t)$ changes from $\Omega$ to $-\Omega$ at $\subt{\tau}{p}$. The dipole blockade is used to drastically reduce the size of the Hilbert space as described in detail in \cite{Wei08}. The simulations are performed for $N = 44 \ldots 54$, $C_6/V^2 = 0.01$, $\Omega \subt{\tau}{p} = 0.32 \ldots 0.74$. The resulting echo curves have been fitted the same way as the experimental data with a parabolic curve. The dephasing rate is again obtained by $\subt{\gamma}{d}=-\ln(V)/\tau$.

\subsection{Electromagnetically induced transparency in an interacting Rydberg gas}\label{section:EIT}

EIT occurs in systems with at least three levels coupled by two laser modes \cite{Har90}. If two of the three levels have a long lifetime compared to the third level the system evolves on the timescale of the third level into a dark state. This dark state is a superposition of the two long living states without a contribution of the `radiative' intermediate state. Thus, probing with a weak laser on the transition between one of the long living states and the radiative state results in a narrow transparency window around the resonance of this transition. 

Such a three-level system can be realized within the two-photon excitation scheme into long living Rydberg states. Again the $\ftS (j=1/2,m_J=1/2)$ Rydberg state is used, which has a lifetime of $\sim\unit[100]{\mu s}$. In the following this state is referred to as \rstate . The ground state \gstate\, is the \fiveS$(f=2,m_f=2)$ state, which does not decay on the timescale of our experiments. Finally, the radiative state \estate\, is given by the intermediate \fiveP$(f=3,m_f=3)$ state, which has a decay rate of $\subt{\Gamma}{eg} = \twopit \unit[6]{MHz}$. EIT in such a ladder system involving a Rydberg state has been investigated previously in a thermal vapour of $^{85}$Rb \cite{Moh07} and in a weakly interacting Rydberg gas \cite{Wea08}.

\begin{figure}[ht]
  \includegraphics{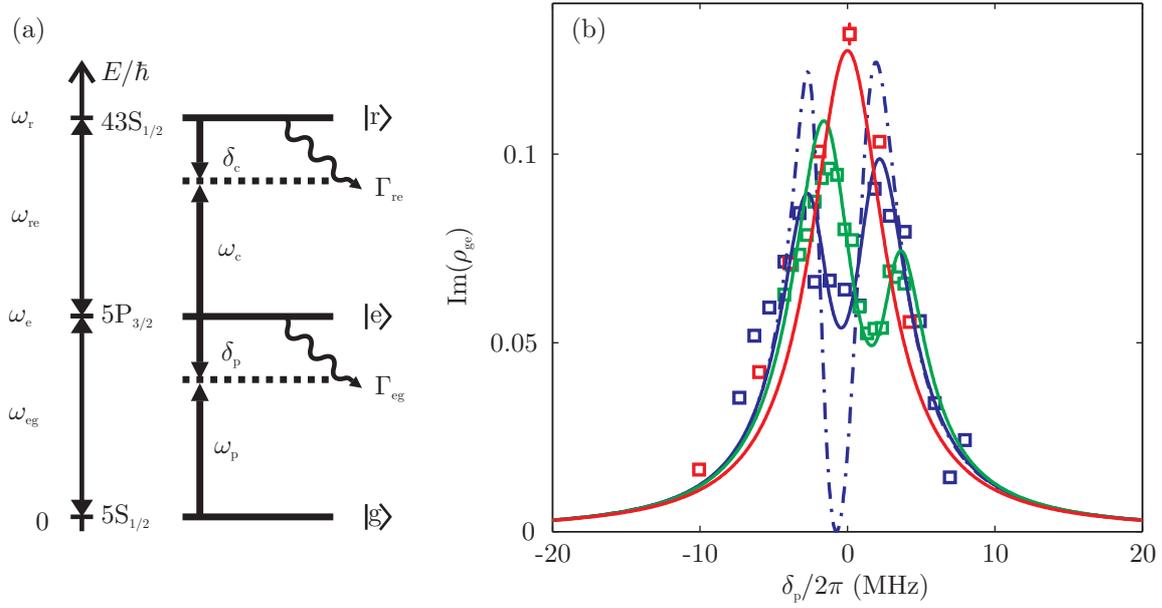}
  \caption{(a) Level scheme of the three-level atom to clarify the symbols used in the text. The states are $\gstate=\fiveS(f=2,m_f=2)$, $\estate=\fiveP(f=3,m_f=3)$ and $\rstate=\ftS(j=1/2,m_j=1/2)$. (b)  Imaginary part of the coherence $\subt{\rho}{ge}$ as a function of the detuning $\subt{\delta}{p}$ of the probe laser. The red curve is taken without the presence of a coupling laser, i.e. $\subt{\Omega}{c}=0$. With the coupling laser on the absorption and, thus, Im($\subt{\rho}{ge}$) decreases for $\subt{\delta}{p}\rightarrow 0$. Im($\subt{\rho}{ge})=0$ is reached if no damping or dephasing occur, i.e. $\subt{\gamma}{rd}=0$ (dashed blue line). With a finite dephasing due to interaction between Rydberg atoms the depth of the dip decreases. The solid lines are calculated by numerically diagonalizing \eqref{eq:Lindblad_01} with $\subt{\Omega}{p}=\unit[\twopit 800]{kHz}$, $\subt{\Omega}{c}=\unit[\twopit 4.5]{MHz}$, $\subt{\gamma}{ed}=0$. The dephasing is $\subt{\gamma}{rd}=\unit[\twopit 3.0]{MHz}$ and $\subt{\gamma}{rd}=\unit[\twopit 3.5]{MHz}$ for the blue and the green data set, respectively. The natural linewidth of \estate\, is $\subt{\Gamma}{eg}=\unit[\twopit 6]{MHz}$ and the \ftS\, Rydberg state \rstate\, has a decay rate of $\subt{\Gamma}{re}=\unit[\twopit 1.6]{kHz}$. Both blue lines are calculated with a detuning $\subt{\delta}{c}=\unit[\twopit 0.75]{MHz}$ and the green line is calculated with $\subt{\delta}{c}=-\unit[\twopit 2.0]{MHz}$. The atomic peak density of ground state atoms is $\subt{n}{g} = \unit[2.4\times 10^{17}]{m^{-3}}$.}
  \label{fig:rotesEIT}
\end{figure}

The system can be described using the Lindblad equation of motion for the density matrix $\vec{\rho}$
\begin{eqnarray}
\label{eq:Lindblad_01}
\dot{\vec{\rho}} = -\frac{i}{\hbar}[\mathcal{H},\vec{\rho}] + \mathcal{L}\, . 
\end{eqnarray}
With the rotating wave approximation, the Hamilton operator reads in the basis $\gstate=(1,0,0)^{\textnormal{\footnotesize t}}$, $\estate=(0,1,0)^{\textnormal{\footnotesize t}}$ and  $\rstate=(0,0,1)^{\textnormal{\footnotesize t}}$
\begin{eqnarray}
\label{eq:Lindblad_02}
\mathcal{H}  = \frac{\hbar}{2}
\renewcommand{\arraystretch}{1.25}
\left(
\begin{array}{ccc}
 0 & \subt{\Omega}{p} & 0 \\
 \subt{\Omega}{p}^{\ast} & -2\subt{\delta}{p} & \subt{\Omega}{c} \\
0 & \subt{\Omega}{c}^{\ast} & -2\delta
\end{array}\right)\, ,
\end{eqnarray}
with the Rabi frequency defined as $\hbar\Omega_{mn}=-E_0d_{mn}$, where $E_0$ is the electric field amplitude and $d_{mn}=\bra{m}e_0r\ket{n}$ is the matrix element for the dipole transition. The detunings are (see \fref{fig:rotesEIT} (a)) $\subt{\delta}{p}=\subt{\omega}{p}-\subt{\omega}{eg}$ and $\subt{\delta}{c}=\subt{\omega}{c}-\subt{\omega}{re}$ for the probe and the coupling laser, respectively. The two-photon detuning is $\delta=\subt{\delta}{p}+\subt{\delta}{c}$.

The Liouville operator $\mathcal{L}$ in \fref{eq:Lindblad_01} 
\begin{eqnarray}
\label{eq:Lindblad_03}
\mathcal{L} = 
\renewcommand{\arraystretch}{1.25}
\left(
\begin{array}{ccc}
 \subt{\Gamma}{eg} \subt{\rho}{ee} & -\frac{1}{2} \subt{\Gamma}{e} \subt{\rho}{ge} & -\frac{1}{2} \subt{\Gamma}{r} \subt{\rho}{gr}\\
 -\frac{1}{2} \subt{\Gamma}{e}\subt{\rho}{eg} &
-\subt{\Gamma}{eg}\subt{\rho}{ee} + \subt{\Gamma}{re}\subt{\rho}{rr} & -\frac{1}{2} (\subt{\Gamma}{e}+\subt{\Gamma}{r})\subt{\rho}{er} \\
 -\frac{1}{2} \subt{\Gamma}{r}\subt{\rho}{rg}  & -\frac{1}{2} (\subt{\Gamma}{e}+\subt{\Gamma}{r})\subt{\rho}{re} & -\subt{\Gamma}{re}\subt{\rho}{rr}
\end{array}\right)\, ,
\renewcommand{\arraystretch}{1}
\end{eqnarray}
refers to the dissipation and dephasing in the three-level atom. The decay rates are $\subt{\Gamma}{e} = \subt{\gamma}{ed}+\subt{\Gamma}{eg}$ and $\subt{\Gamma}{r}=\subt{\gamma}{rd}+\subt{\Gamma}{re}$, with $\subt{\Gamma}{eg}$ and $\subt{\Gamma}{re}$ being the natural linewidth of \estate\, and \rstate . In addition analogue to \cite{Fle05} the dephasing rates of these states are taken into account. The dephasing rates are denoted by $\subt{\gamma}{ed}$ and $\subt{\gamma}{rd}$.

The dominant contribution to the dephasing is due to the interaction between Rydberg atoms, which can be assumed to be much larger than any other dephasing. Thus, in the following the dephasing rate of the intermediate state $\subt{\gamma}{ed}$ is neglected.
 
The two-photon excitation scheme to measure the dephasing rate $\subt{\gamma}{rd}$ is mostly the same as for the rotary echo in the ODT. However, the direction for the probe laser driving the \fiveS$(f=2,m_f=+2)$ to \fiveP$(f=3,m_f=+3)$ excitation is now the $y$-axis (see \fref{fig:Setup} (b)), which is also the quantization axis. This beam has a diameter of \unit[13.5]{mm} and is detected by a CCD camera. The \unit[480]{nm} coupling laser is tuned to the \fiveP$(f=3,m_f=+3)$ to \ftS$(j=1/2,m_j=+1/2)$ resonance and linearly polarized along the $x$-direction, while still travelling along $-z$. The Rabi frequency is $\subt{\Omega}{p}=\unit[\twopit 2]{MHz}$ and $\subt{\Omega}{c}=\unit[\twopit 8]{MHz}$ for the \unit[780]{nm} and \unit[480]{nm} laser, respectively. The Rabi frequency for the coupling transition is obtained by calculating the dipole matrix element between the \fiveP\, and \ftS\, states. The calculations are in good agreement with \cite{Joh08}.  

In order to decrease the optical density of the atomic sample a Landau-Zener sweep and a time of flight of \unit[100]{$\mu$s} is used. This lowers the peak density to values between $\subt{n}{g}\sim \unit[2\times 10^{17}]{m^{-3}}$ and $\unit[5\times 10^{17}]{m^{-3}}$ at a temperature of $\unit[6.2]{\mu K}$. The cloud is excited for \unit[100]{$\mu$s} after which the Rydberg atoms are field ionized and detected by the MCP. Note that during illumination the density of the cloud is only decreasing by $\unit[\sim\! 50]{\%}$. 

In steady state the coherence $\subt{\rho}{ge}$ is obtained from the absorption images in the following way. The scattering cross section is $\sigma=\sigma_0\subt{\Gamma}{eg}\textnormal{Im}(\subt{\rho}{ge})/\subt{\Omega}{p}$, with the on-resonance scattering cross section $\sigma_0=\sigma(\subt{\delta}{p} + \subt{\delta}{c}=0)$. The atom number obtained from an off-resonant absorption image is $N(\delta)\sim \textrm{OD}(\delta)/\sigma_0$, where $\textrm{OD}(\delta)$ is the optical density depending on the two-photon detuning.
%
%
%The fit routines for obtaining the atom number assume that the scattering cross section is constantly $\sigma_0$. Thus, we obtain a `virtual' atom number $N(\delta)$ depending on the two-photon detuning from the cloud fits. The real atom number is found for $\delta=0$ and is denoted by $N_0$. 
Since $\sigma/\sigma_0 = N(\delta)/N_0$, the imaginary part of the coherence $\subt{\rho}{ge}$ in steady state is found to be
\begin{eqnarray}
\label{eq:Imrhoeg}
\textnormal{Im}(\subt{\rho}{ge}) = \frac{\subt{\Omega}{p}}{\subt{\Gamma}{eg}}\cdot\frac{N(\delta)}{N_0}\, .
\end{eqnarray}
In the case of electromagnetically induced transparency a certain number of the ground state atoms are excited to the Rydberg state and do not contribute to the coherence $\subt{\rho}{ge}$. Equation \eqref{eq:Imrhoeg} must be corrected by this atom number and reads
\begin{eqnarray}
\label{eq:Imrhoeg_2}
\textnormal{Im}(\subt{\rho}{ge}) = \frac{\subt{\Omega}{p}}{\subt{\Gamma}{eg}}\cdot\frac{N(\delta)}{N_0 - \textrm{max}(\subt{N}{R})}\, ,
\end{eqnarray}
where $\textrm{max}(\subt{N}{R})$ being the Rydberg atom on two-photon resonance.

On the other hand Im($\subt{\rho}{ge}$) can be calculated by solving \eqref{eq:Lindblad_01} in steady state, i.e. $\dot{\vec{\rho}}=0$. \Fref{fig:rotesEIT} (b) shows three measurements of Im($\subt{\rho}{ge}$) with $\subt{\Omega}{c}=0$ (red) and $\subt{\Omega}{c}=\unit[\twopit 4.5]{MHz}$ (green and blue lines). In the presence of the coupling laser a clear signature for the population of the dark state is visible, resulting in a decrease of the absorption for $\subt{\delta}{p}\rightarrow 0$. 

Without dephasing, i.e. $\subt{\gamma}{rd} = \subt{\gamma}{ed} = 0$, the absorption would tend to zero, shown as a dashed blue line in \fref{fig:rotesEIT} (b). However, in the presence of interaction between Rydberg atoms the dephasing $\subt{\gamma}{rd}$ of the Rydberg level is finite. The curves in \fref{fig:rotesEIT} (b) are calculated with $\subt{\gamma}{rd}=\unit[\twopit 3.0]{MHz}$ and $\subt{\gamma}{rd}=\unit[\twopit 3.5]{MHz}$ for the blue and the green data set, respectively. The measurements were taken with a detuning of $\subt{\delta}{c}=\unit[\twopit 0.75]{MHz}$ and $\subt{\delta}{c}=-\unit[\twopit 2.0]{MHz}$ for the blue and the green data set, respectively. The feature has a linewidth smaller than the natural linewidth of the $\estate$ to $\gstate$ transition and, hence, is a direct proof of the coherent population of the dark state.

\begin{figure}[ht]
  \includegraphics{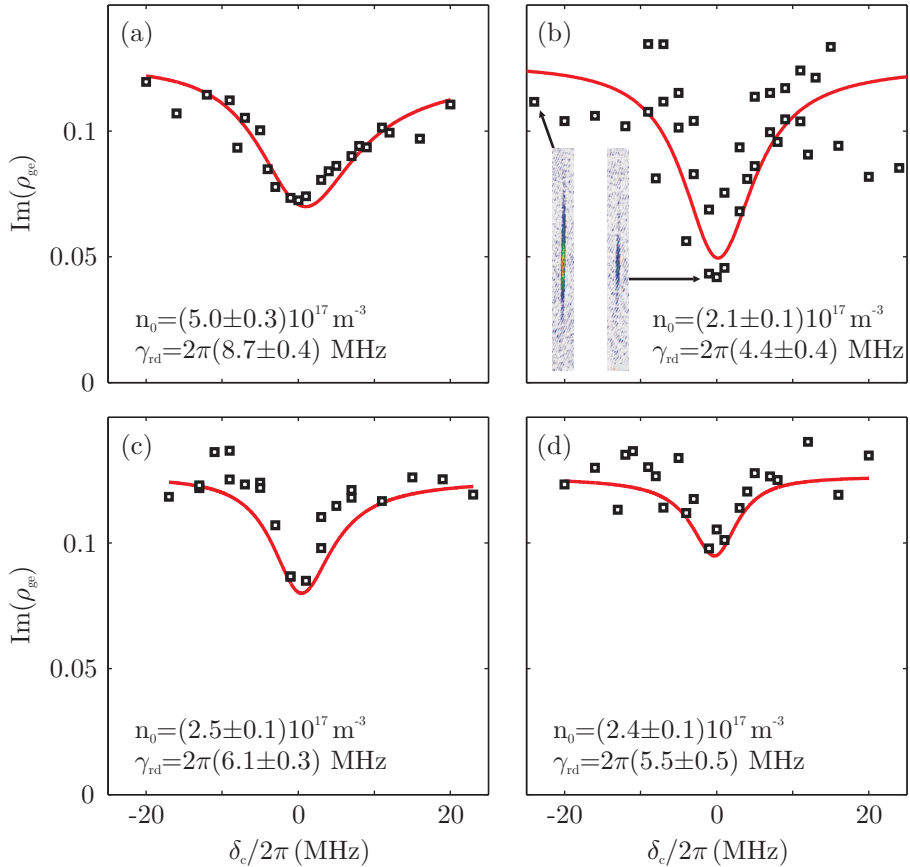}
  \caption{Imaginary part of $\subt{\rho}{ge}$ as a function of the detuning $\subt{\delta}{c}$ of the coupling laser for different peak densities $\subt{n}{g}$ of ground state atoms and Rabi frequencies $\subt{\Omega}{c}$. The data are normalized for $|\subt{\delta}{c}|\gg 0$ to the steady state value of Im($\subt{\rho}{eg}$) for $\subt{\Omega}{c}=0$. The red lines show the numerically obtained solution of (\ref{eq:Lindblad_01}) with a probe Rabi frequency $\subt{\Omega}{p}=\unit[\twopit 800]{kHz}$. The coupling Rabi frequency in figures (a) and (b) are $\subt{\Omega}{c}=\unit[\twopit 6.5]{MHz}$, in figure (c) $\subt{\Omega}{c}=\unit[\twopit 4.6]{MHz}$ and in figure (d) $\subt{\Omega}{c}=\unit[\twopit 3.3]{MHz}$. The insets in figure (b) show two typical absorption pictures of the atomic cloud for $\subt{\delta}{c}\neq 0$ (left) and $\subt{\delta}{c}=0$ (right). The red and grey colour in these pictures corresponds to an optical density of 0.4 and 0, respectively.}
  \label{fig:blauesEIT}
\end{figure}

The dephasing rates can be measured by scanning the coupling instead of scanning the probe laser with the advantage of removing the absorption line shape from the data. The measurements of Im($\subt{\rho}{ge}$) as a function of $\subt{\delta}{c}$ are shown in \fref{fig:blauesEIT} for different peak densities of ground state atoms $\subt{n}{g}$ and different coupling Rabi frequencies $\subt{\Omega}{c}$. Again, approaching the two-photon resonance a reduction of Im($\subt{\rho}{ge}$) is found. A fit function to obtain the dephasing rates from the data shown in \fref{fig:blauesEIT} can be found by solving \eqref{eq:Lindblad_01} in the steady state. In case of a perturbative probe laser it is sufficient to take only the first order expansion in the probe Rabi frequency \cite{Fle05}
\begin{eqnarray}
\label{eq:Fit}
\textnormal{Im}(\subt{\rho}{ge}) \propto \frac{4\delta^2\subt{\Gamma}{e} + \subt{\Gamma}{r}\left(|\subt{\Omega}{c}|^2 + \subt{\Gamma}{e}\subt{\Gamma}{r}\right)}{||\subt{\Omega}{c}|^2 + \left(\subt{\Gamma}{e} - 2i\subt{\delta}{p}\right)\left(\subt{\Gamma}{r} - 2i\delta\right)|^2} \, .
\end{eqnarray}
This function is used to pre-fit the data shown in \fref{fig:blauesEIT}. The only free fit parameters in this function are the two-photon detuning $\delta$ and the rate $\subt{\Gamma}{r} = \subt{\Gamma}{re} + \subt{\gamma}{rd}\simeq\subt{\gamma}{rd}$. The value for the maximal coupling Rabi frequency is $\subt{\Omega}{c}=\twopit\unit[6.5]{MHz}$ and the rate $\subt{\Gamma}{e} = \subt{\Gamma}{eg} + \subt{\gamma}{ed}\simeq\subt{\Gamma}{eg}$. The parameters $\subt{\Gamma}{e}=\subt{\Gamma}{eg}$ and $\subt{\Omega}{c}$ were kept fix. In order to take the finite probe Rabi frequency of $\subt{\Omega}{p}=\unit[\twopit 800]{kHz}$ the imaginary part of the coherence is afterwards calculated by numerical solving (\ref{eq:Lindblad_01}). The red curves in \fref{fig:blauesEIT} show the result of these calculations.

\section{Results and discussion}\label{section:ResultsAndDiscussion}

\begin{figure}[ht]
  \includegraphics{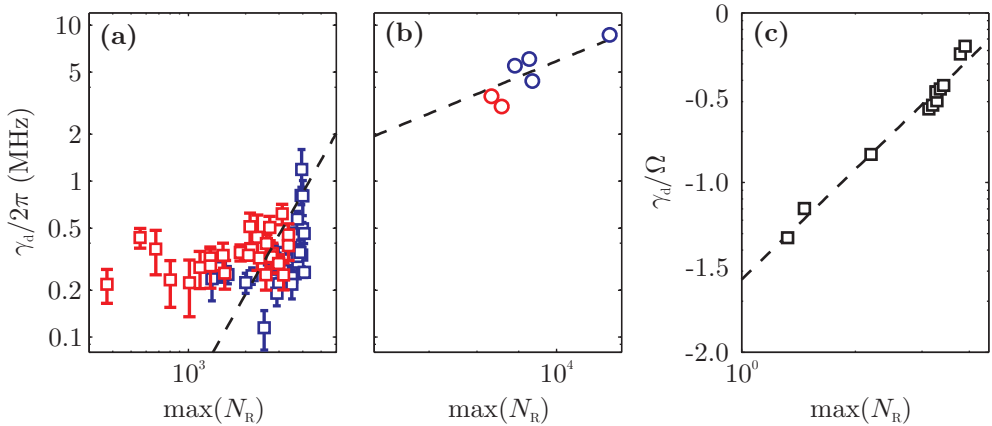}
  \caption{Collection of the dephasing rates $\subt{\gamma}{d}$ as a function of the measured maximal Rydberg atom number $\subt{N}{R}$. All figures are plotted double logarithmic. Figure (a) shows the results from the rotary echo measurements in the magnetic trap (\textcolor{blue}{\pmb{$\square$}}) and in the ODT (\textcolor{red}{\pmb{$\square$}}). Figure (b) shows the dephasing rates $\subt{\gamma}{d}=\subt{\gamma}{rd}$ obtained from the EIT measurements presented in figure \ref{fig:rotesEIT} (\textcolor{red}{\pmb{$\bigcirc$}}) and figure \ref{fig:blauesEIT} (\textcolor{blue}{\pmb{$\bigcirc$}}). Theoretically investigated dephasing rates for numerically calculated rotary echo experiments are shown in figure (c). The dashed line in figure (c) is a fit to the theoretical data, whilst the dashed lines in (a) and (b) have the same slope as in (c), but fitted offsets (see text).}
 \label{fig:dephasing}
\end{figure}

The dephasing rates gathered from different experiments are shown in figures \ref{fig:dephasing} (a) and (b) against the measured maximal Rydberg atom number. In the case of the rotary echo data, for the experiment as well as for the simulation, this number is max$(\subt{N}{R})=\subt{N}{R}(\subt{\tau}{p}=0)$. In the EIT experiments max$(\subt{N}{R})$ is the Rydberg atom number on two-photon resonance, i.e. max$(\subt{N}{R})=\subt{N}{R}(\delta=0)$. The data show a tendency that for higher Rydberg atom numbers the dephasing is increasing. This assertion is supported by the results of the numerical calculation. The only dephasing in the simulation is due to the interaction between Rydberg atoms and is not affected by any experimental uncertainties, e.g. laser linewidth.

In order to identify a power law dependence of the dephasing on max$(\subt{N}{R})$ the data is plotted double logarithmic. A power law of the form $\subt{\gamma}{d}=\subt{a}{c}\,\textnormal{max}(\subt{N}{R})^b$ is shown in figure \ref{fig:dephasing} as a dashed line. The fit to the calculated rotary echo data yields $\subt{a}{c}=\unit[(0.03\pm 0.01)]{\Omega}$ and $b=2.16\pm 0.12$. The obtained exponent $b$ reflects the nature of the dephasing of the ultracold sample of Rydberg atoms, namely the van der Waals interaction $\subt{V}{vdW}=C_6 \subt{N}{R}^2$. Power laws with a constant exponent $b=2.16$ are fitted to the data shown in figure \ref{fig:dephasing} (a) and (b). The resulting coefficients are $\subt{a}{a}\simeq \subt{a}{b}=\unit[(0.09\pm 0.02)]{s^{-1}}$. Note that the definition of the dephasing rates are different for both experimental sequences and, thus, they might differ by a constant factor. However, although both dephasing rates are based on entirely different measurements the results are described by the same power law, which is furthermore in good agreement with the theoretical prediction of the rotary echo experiment. It remains to be investigated if the obtained power law for the dephasing rate is a result of an underlying universality similar to the observed evidences for a universal scaling in ultracold Rydberg gases \cite{Low08}.

The dephasing rate due to a frequency uncertainty is estimated to be $\unit[\simeq\! 1.5]{MHz}$ on the minute time scale. From the measurement shown in the inset of figure \fref{fig:expDecay} an upper bound for the laser linewidth on the \unit[100]{ns} timescale of $\unit[\sim\! 200]{kHz}$ can be found. The laser linewidth causes the same effect on the data as the dephasing due to the interaction does, namely is decreases the visibility of the signal. This constant dephasing is observed as a kink in figure \ref{fig:dephasing} (a) for Rydberg atom numbers max$(\subt{N}{R})< 2\times 10^3$. However, the `dephasing' caused by the instrumentation is much smaller than the observed additional dephasing due to the interaction between the Rydberg atoms. This argument is strengthened by the numerical simulation shown in \fref{fig:dephasing} (c), since these calculation do not take any laser linewidth into account. The simulation reflect the same overall behaviour of the dephasing rate as the experimental data, namely an increase of $\subt{\gamma}{d}$ with increasing Rydberg atom number.

The dephasing of the Rydberg state could be tailored by increasing the confinement of the atomic sample in radial direction in order to study the dephasing rate more quantitively in future experiments. If the radial width becomes smaller than the blockade radius, i.e. the situation becomes purely one dimensional, the reduced number of next neighbours reduces the dephasing due to interaction. Conducting the experiments in an ODT comes with the advantage that the dimensionality is already reduced in comparison to the confinement in the magnetic trap. A disadvantage, which will be addressed in future experiments, is the adjustability of the atomic density in the ground state. The Landau-Zener sweep technique to decrease the density in the ODT comprises an additional heating due to the changed number of atoms in the magnetic trap, which are loaded into the ODT. The lowered thermalization rate in the axial direction might be an explanation for this effect.

\section{Conclusion}
To conclude, two different types of experimental techniques, which allow to investigate the dephasing rates in a system of ultracold Rydberg atoms, are presented. The dephasing rates due to the interaction between Rydberg atoms were found to exceed all other dephasing rates, e.g. caused by instrumentation. The tendency of an increase of the dephasing rate with increasing Rydberg atom number is confirmed by conducting numerical calculations for the rotary echo experiment. Both, the measurement as well as the numerical simulations, are well described by the same exponent for a power law dependence of the dephasing rate on the maximal Rydberg atom number and is close to the naively expected one for a dephasing due to the van der Waals interaction. Furthermore, it is shown that the dephasing coefficients of the power law dependence are equal within the accuracy of the measurement. This result is remarkable in the sense, that the two experiments that are used to obtain the dephasing rates are conducted on completely different time scales. In future experiments strongly interacting frozen Rydberg gases will serve as a test ground for decoherence theories, where the dimensionality, the strength and the nature of the interaction can be tailored.

We acknowledge financial support from the Deutsche Forschungsgemeinschaft within the SFB/TRR21 and under the contract PF 381/4-1. U.R. acknowledges support from the Landesgraduiertenf\"{o}rderung Baden-W\"{u}rttemberg.

%\bibliographystyle{unsrt}
%\bibliography{Literature}

\section{References}
\begin{thebibliography}{10}

\bibitem{Leg87}
A.~J. Leggett, S.~Chakravarty, A.~T. Dorsey, Matthew P.~A. Fisher, Anupam Garg,
  and W.~Zwerger.
\newblock Dynamics of the dissipative two-state system.
\newblock {\em Rev. Mod. Phys.}, 59(1):1--85, Jan 1987.

\bibitem{Weiss08}
Ulrich Weiss.
\newblock {\em Quantum dissipative systems}.
\newblock World Scientific Publishing, 2008.

\bibitem{Jak00}
D.~Jaksch, J.~I. Cirac, P.~Zoller, S.~L. Rolston, R.~C\^ot\'e, and M.~D. Lukin.
\newblock Fast quantum gates for neutral atoms.
\newblock {\em Phys. Rev. Lett.}, 85(10):2208--2211, September 2000.

\bibitem{Luk01}
M.~D. {Lukin}, M.~{Fleischhauer}, R.~{Cote}, L.~M. {Duan}, D.~{Jaksch}, J.~I.
  {Cirac}, and P.~{Zoller}.
\newblock {Dipole Blockade and Quantum Information Processing in Mesoscopic
  Atomic Ensembles}.
\newblock {\em Phys. Rev. Lett.}, 87(3):037901, July 2001.

\bibitem{Sin04}
Kilian Singer, Markus Reetz-Lamour, Thomas Amthor, Luis~Gustavo Marcassa, and
  Matthias Weidem\"{u}ller.
\newblock Suppression of excitation and spectral broadening induced by
  interactions in a cold gas of rydberg atoms.
\newblock {\em Phys. Rev. Lett.}, 93(16):163001, October 2004.

\bibitem{Ton04}
D.~Tong, S.~M. Farooqi, J.~Stanojevic, S.~Krishnan, Y.~P. Zhang,
  R.~C\^{o}t\'{e}, E.~E. Eyler, and P.~L. Gould.
\newblock Local blockade of rydberg excitation in an ultracold gas.
\newblock {\em Phys. Rev. Lett.}, 93(6):063001, August 2004.

\bibitem{Cub05a}
T.~{Cubel Liebisch}, A.~{Reinhard}, P.~R. {Berman}, and G.~{Raithel}.
\newblock {Atom Counting Statistics in Ensembles of Interacting Rydberg Atoms}.
\newblock {\em Phys. Rev. Lett.}, 95(25):253002, December 2005.

\bibitem{Vog06}
T.~{Vogt}, M.~{Viteau}, J.~{Zhao}, A.~{Chotia}, D.~{Comparat}, and P.~{Pillet}.
\newblock {Dipole Blockade at F{\"o}rster Resonances in High Resolution Laser
  Excitation of Rydberg States of Cesium Atoms}.
\newblock {\em Phys. Rev. Lett.}, 97(8):083003, August 2006.

\bibitem{Hei07}
R.~Heidemann, U.~Raitzsch, V.~Bendkowsky, B.~Butscher, R.~L\"{o}w, L.~Santos,
  and T.~Pfau.
\newblock Evidence for coherent collective rydberg excitation in the strong
  blockade regime.
\newblock {\em Phys. Rev. Lett.}, 99(16):163601, 2007.

\bibitem{Moh07}
A.~K. {Mohapatra}, T.~R. {Jackson}, and C.~S. {Adams}.
\newblock {Coherent Optical Detection of Highly Excited Rydberg States Using
  Electromagnetically Induced Transparency}.
\newblock {\em Phys. Rev. Lett.}, 98(11):113003, March 2007.

\bibitem{Rai08}
U.~Raitzsch, V.~Bendkowsky, R.~Heidemann, B.~Butscher, R.~L\"{o}w, and T.~Pfau.
\newblock {Echo Experiments in a Strongly Interacting Rydberg Gas}.
\newblock {\em Phys. Rev. Lett.}, 100(1):013002, January 2008.

\bibitem{Wea08}
K~J Weatherill, J~D Pritchard, R~P Abel, M~G Bason, A~K Mohapatra, and C~S
  Adams.
\newblock Electromagnetically induced transparency of an interacting cold
  rydberg ensemble.
\newblock {\em Journal of Physics B: Atomic, Molecular and Optical Physics},
  41(20):201002, October 2008.

\bibitem{Sol59}
I.~{Solomon}.
\newblock {Rotary Spin Echoes}.
\newblock {\em Phys. Rev. Lett.}, 2:301--302, April 1959.

\bibitem{Har90}
S.~E. {Harris}, J.~E. {Field}, and A.~{Imamo{\u g}lu}.
\newblock {Nonlinear optical processes using electromagnetically induced
  transparency}.
\newblock {\em Phys. Rev. Lett.}, 64:1107--1110, March 1990.

\bibitem{Bol91}
K.-J. Boller, A.~Imamo\u{g}lu, and S.~E. Harris.
\newblock Observation of electromagnetically induced transparency.
\newblock {\em Phys. Rev. Lett.}, 66(20):2593--2596, May 1991.

\bibitem{Loe07}
R.~L{\"o}w, U.~{Raitzsch}, R.~Heidemann, V.~Bendkowsky, B.~Butscher,
  A.~Grabowski, and T.~Pfau.
\newblock Apparatus for excitation and detection of rydberg atoms in quantum
  gases.
\newblock \emph{Preprint} arXiv:0706.2639v1, 2007.

\bibitem{Zen32}
Clarence Zener.
\newblock Non-adiabatic crossing of energy levels.
\newblock {\em Proceedings of the Royal Society of London. Series A},
  137(833):696--702, September 1932.

\bibitem{Wei08}
H.~Weimer, R.~L\"{o}w, T.~Pfau, and H.~P. B\"{u}chler.
\newblock Quantum critical behavior in stronlgy interacting rydberg gases.
\newblock \emph{Accepted for publication in Phys. Rev. Lett.}, June 2008.

\bibitem{Fle05}
M.~{Fleischhauer}, A.~{Imamoglu}, and J.~P. {Marangos}.
\newblock {Electromagnetically induced transparency: Optics in coherent media}.
\newblock {\em Rev. Mod. Phys.}, 77:633--673, July 2005.

\bibitem{Joh08}
T.~A. Johnson, E.~Urban, T.~Henage, L.~Isenhower, D.~D. Yavuz, T.~G. Walker,
  and M.~Saffman.
\newblock Rabi oscillations between ground and rydberg states with
  dipole-dipole atomic interactions.
\newblock {\em Phys. Rev. Lett.}, 100(11):113003, 2008.

\bibitem{Low08}
R.~L\"{o}w, H.~Weimer, U.~{Raitzsch}, R.~{Heidemann}, V.~{Bendkowsky}, H.~P.
  B\"{u}chler, and T.~{Pfau}.
\newblock Universal scaling behavior in a strongly interacting rydberg gas.
\newblock In preparation, 2008.

\end{thebibliography}

\end{document}